\documentclass[12pt,a4paper]{article}
\usepackage{amsmath}
\usepackage{bm}
\usepackage{amsfonts}
\usepackage{graphicx}
\usepackage[normal]{caption2}
\usepackage{subfigure}
\usepackage{rotating}
\usepackage{citesort}
\usepackage{lscape}
\usepackage{section}
\begin{document}
\renewcommand\arraystretch{1.1}
\setlength{\abovecaptionskip}{0.1cm}
\setlength{\belowcaptionskip}{0.5cm}
\pagestyle{empty}
\newpage
\pagestyle{plain} \setcounter{page}{1} \setcounter{lofdepth}{2}
\begin{center} {\large\bf Study of fragmentation and momentum correlations in heavy-ion collisions}\\
\vspace*{0.4cm}
{\bf Sakshi Gautam}\footnote{Email:~sakshigautm@gmail.com} {and \bf Rajni Kant}\\
{\it  Department of Physics, Panjab University, Chandigarh -160
014, India.\\}
\end{center}
The role of momentum correlations is studied in the production of
light and medium mass fragments by imposing momentum cut in
clusterization the phase space. Our detailed investigation shows
that momentum cut has major role to play in the emission of
fragments.
\newpage
\baselineskip 20pt
\section{Introduction}
 \par
It is well known that colliding nuclei break into several small
and medium size pieces and a lot of nucleons are also emitted.
This subfield, known as multifragmentation, gained momentum after
several theoretical and experimental groups around the world put
their collective efforts to understand this process \cite{bege}.
Because of accumulation of experimental data on
multifragmentation, one has the opportunity to study the role of
dynamical correlations in fragment formation. Theoretically, the
availability of large number of models makes the situation worse
\cite{bege,aich1,qmd1,dorso}. Due to fact that fragmentation need
fluctuations and correlations, the molecular dynamical models are
the only resource in theoretical domains. The molecular dynamics
(n-body) approach is well suited as it incorporates the
correlations and fluctuations among the nucleons. We will,
therefore, use quantum molecular dynamics (QMD) model
\cite{aich1,qmd1} to study the dynamics of heavy-ion collisions.
Since every model simulate single nucleon, one needs to have
afterburner to clusterize the phase space. In a very simple
picture, we can define a cluster by using space correlations. This
method is known as minimum spanning tree (MST) method
\cite{jsingh}. In this method, we allow nucleons to form a cluster
if their centroids are less than 4 fm. This method works fine when
the system is very dilute. At the same time fragments formed in
MST method will be highly unstable (especially in central
collisions) as there the two nucleons may not be well formed and
therefore can be unstable that will decay after a while. In order
to filter out such unstable fragments, we impose another cut in
terms of relative momentum of nucleons. This method, dubbed as
minimum spanning tree with momentum cut (MSTP) method was
discussed by Puri \emph{et al.} \cite{kumar1}. Unfortunately this
study was restricted to heavier systems like $^{93}$Nb+$^{93}$Nb
and $^{197}$Au+$^{197}$Au reactions. The role of momentum cut on
the fragment structure of lighter systems is still unclear. We aim
to address this in present paper.
\par
Exclusively we plan to\\
 (i) see the
role of momentum cut on the fragment structure of lighter colliding systems.\\
(ii) and see the role of colliding geometry on the fragment
structure with momentum cut being imposed. \\
The present study is carried out within the framework of QMD model
\cite{aich1,qmd1} which is described in the following section.

\par
\section{The Formalism}
\subsection{Quantum Molecular dynamics (QMD) model}
\par
We describe the time evolution of a heavy-ion reaction within the
framework of Quantum Molecular Dynamics (QMD) model
\cite{aich1,qmd1} which is based on a molecular dynamics picture.
This model has been successful in explaining collective flow
\cite{sood2}, elliptic flow \cite{kumar3}, multifragmentation
\cite{dhawan} as well as dense and hot matter \cite{fuchs}. Here
each nucleon is represented by a coherent state of the form
\begin{equation}
\phi_{\alpha}(x_1,t)=\left({\frac {2}{L \pi}}\right)^{\frac
{3}{4}} e^{-(x_1-x_{\alpha }(t))^2}
e^{ip_{\alpha}(x_1-x_{\alpha})} e^{-\frac {i p_{\alpha}^2 t}{2m}}.
\label {e1}
\end{equation}
Thus, the wave function has two time dependent parameters
$x_{\alpha}$ and $p_{\alpha}$.  The total n-body wave function is
assumed to be a direct product of coherent states:
\begin{equation}
\phi=\phi_{\alpha}
(x_1,x_{\alpha},p_{\alpha},t)\phi_{\beta}(x_2,x_{\beta},
p_{\beta},t)....,         \label {e2}
\end{equation}
where antisymmetrization is neglected. One should, however, keep
in the mind that the Pauli principle, which is very important at
low incident energies, has been taken into account. The initial
values of the parameters are chosen in a way that the ensemble
($A_T$+$A_P$) nucleons give a proper density distribution as well
as a proper momentum distribution of the projectile and target
nuclei. The time evolution of the system is calculated using the
generalized variational principle. We start out from the action
\begin{equation}
S=\int_{t_1}^{t_2} {\cal {L}} [\phi,\phi^{*}] d\tau, \label {e3}
\end{equation}
with the Lagrange functional
\begin{equation}
{\cal {L}} =\left(\phi\left|i\hbar \frac
{d}{dt}-H\right|\phi\right), \label {e4}
\end{equation}
where the total time derivative includes the derivatives with
respect to the parameters. The time evolution is obtained by the
requirement that the action is stationary under the allowed
variation of the wave function
\begin{equation}
\delta S=\delta \int_{t_1}^{t_2} {\cal {L}} [\phi ,\phi^{*}] dt=0.
\label{e5}
\end{equation}
If the true solution of the Schr\"odinger equation is contained in
the restricted set of wave function
$\phi_{\alpha}\left({x_{1},x_{\alpha},p_{\alpha}}\right),$ this
variation of the action gives the exact solution of the
Schr\"odinger equation. If the parameter space is too restricted,
we obtain that wave function in the restricted parameter space
which comes close to the solution of the Schr\"odinger equation.
Performing the variation with the test wave function (2), we
obtain for each parameter $\lambda$ an Euler-Lagrange equation;
\begin{equation}
\frac{d}{dt} \frac{\partial {\cal {L}}}{\partial {\dot
{\lambda}}}-\frac{\partial \cal {L}} {\partial \lambda}=0.
\label{e6}
\end{equation}
For each coherent state and a Hamiltonian of the form, \\

$H=\sum_{\alpha}
\left[T_{\alpha}+{\frac{1}{2}}\sum_{\alpha\beta}V_{\alpha\beta}\right]$,
the Lagrangian and the Euler-Lagrange function can be easily
calculated \cite{aich2}
\begin{equation}
{\cal {L}} = \sum_{\alpha}{\dot {\bf x}_{\alpha}} {\bf
p}_{\alpha}-\sum_{\beta} \langle{V_{\alpha
\beta}}\rangle-\frac{3}{2Lm}, \label{e7}
\end{equation}
\begin{equation}
{\dot {\bf x}_{\alpha}}=\frac{{\bf
p}_\alpha}{m}+\nabla_{p_{\alpha}}\sum_{\beta} \langle{V_{\alpha
\beta}}\rangle, \label {e8}
\end{equation}
\begin{equation}
{\dot {\bf p}_{\alpha}}=-\nabla_{{\bf x}_{\alpha}}\sum_{\beta}
\langle{V_{\alpha \beta}}\rangle. \label {e9}
\end{equation}
Thus, the variational approach has reduced the n-body
Schr\"odinger equation to a set of 6n-different equations for the
parameters which can be solved numerically. If one inspects  the
formalism carefully, one finds that the interaction potential
which is actually the Br\"{u}ckner G-matrix can be divided into
two parts: (i) a real part and (ii) an imaginary part. The real
part of the potential acts like a potential whereas imaginary part
is proportional to the cross section.

In the present model, interaction potential comprises of the
following terms:
\begin{equation}
V_{\alpha\beta} = V_{loc}^{2} + V_{loc}^{3} + V_{Coul} + V_{Yuk}
 \label {e10}
 \end {equation}

$V_{loc}$ is the Skyrme force whereas $V_{Coul}$, $V_{Yuk}$ and
$V_{MDI}$ define, respectively, the Coulomb, and Yukawa
potentials. The Yukawa term separates the surface which also plays
the role in low energy processes like fusion and cluster
radioactivity \cite{puri}. The expectation value of these
potentials is calculated as
\begin{eqnarray}
V^2_{loc}& =& \int f_{\alpha} ({\bf p}_{\alpha}, {\bf r}_{\alpha},
t) f_{\beta}({\bf p}_{\beta}, {\bf r}_{\beta}, t)V_I ^{(2)}({\bf
r}_{\alpha}, {\bf r}_{\beta})
\nonumber\\
&  & \times {d^{3} {\bf r}_{\alpha} d^{3} {\bf r}_{\beta}
d^{3}{\bf p}_{\alpha}  d^{3}{\bf p}_{\beta},}
\end{eqnarray}
\begin{eqnarray}
V^3_{loc}& =& \int  f_{\alpha} ({\bf p}_{\alpha}, {\bf
r}_{\alpha}, t) f_{\beta}({\bf p}_{\beta}, {\bf r}_{\beta},t)
f_{\gamma} ({\bf p}_{\gamma}, {\bf r}_{\gamma}, t)
\nonumber\\
&  & \times  V_I^{(3)} ({\bf r}_{\alpha},{\bf r}_{\beta},{\bf
r}_{\gamma}) d^{3} {\bf r}_{\alpha} d^{3} {\bf r}_{\beta} d^{3}
{\bf r}_{\gamma}
\nonumber\\
&  & \times d^{3} {\bf p}_{\alpha}d^{3} {\bf p}_{\beta} d^{3} {\bf
p}_{\gamma}.
\end{eqnarray}
where $f_{\alpha}({\bf p}_{\alpha}, {\bf r}_{\alpha}, t)$ is the
Wigner density which corresponds to the wave functions (eq. 2). If
we deal with the local Skyrme force only, we get
{\begin{equation} V^{Skyrme} = \sum_{{\alpha}=1}^{A_T+A_P}
\left[\frac {A}{2} \sum_{{\beta}=1} \left(\frac
{\tilde{\rho}_{\alpha \beta}}{\rho_0}\right) + \frac
{B}{C+1}\sum_{{\beta}\ne {\alpha}} \left(\frac {\tilde
{\rho}_{\alpha \beta}} {\rho_0}\right)^C\right].
\end{equation}}

Here A, B and C are the Skyrme parameters which are defined
according to the ground state properties of a nucleus. Different
values of C lead to different equations of state. A larger value
of C (= 380 MeV) is often dubbed as stiff equation of state.The
finite range Yukawa ($V_{Yuk}$) and effective Coulomb potential
($V_{Coul}$) read as:
\begin{equation}
V_{Yuk} = \sum_{j, i\neq j} t_{3}
\frac{exp\{-|\textbf{r}_{\textbf{i}}-\textbf{r}_{\textbf{j}}|\}/\mu}{|\textbf{r}_{\textbf{i}}-\textbf{r}_{\textbf{j}}|/\mu},
\end{equation}
\begin{equation}
V_{Coul} = \sum_{j, i\neq
j}\frac{Z_{eff}^{2}e^{2}}{|\textbf{r}_{\textbf{i}}-\textbf{r}_{\textbf{j}}|}.
\end{equation}
\par
The Yukawa interaction (with $t_{3}$= -6.66 MeV and $\mu$ = 1.5
fm) is essential for the surface effects. The relativistic effect
does not play role in low incident energy of present interest
\cite{lehm}.
\par
The phase space of nucleons is stored at several time steps. The
QMD model does not give any information about the fragments
observed at the final stage of the reaction. In order to construct
 the fragments, one needs
clusterization algorithms. We shall concentrate here on the MST
and MSTP methods.
\par
 According to MST method
\cite{jsingh}, two nucleons are allowed to share the same fragment
if their centroids are closer than a distance $r_{min}$,
\begin{equation}
|\textbf{r}_{\textbf{i}}-\textbf{r}_{\textbf{j}}| \leq r_{min}.
\end{equation}
where $\textbf{r}_{\textbf{i}}$ and $\textbf{r}_{\textbf{j}}$ are
the spatial positions of both nucleons and r$_{min}$ taken to be
4fm.
\par
 For MSTP method,we impose a additional cut in the
momentum space, i.e., we allow only those nucleons to form a
fragment which in addition to equation(16) also satisfy
\begin{eqnarray}
|\textbf{p}_{\textbf{i}}-\textbf{p}_{\textbf{j}}| \leq p_{min},
\end{eqnarray}
where p$_{min}$ = 150 MeV/c.
\par
\section{Results and Discussion}
We simulated the reactions of $^{12}$C+$^{12}$C ,
$^{40}$Ca+$^{40}$Ca, $^{96}$Zr+$^{96}$Zr  and
$^{197}$Au+$^{197}$Au at 100 and 400 MeV/nucleon at $\hat{b}$ =
0.0, 0.2, 0.4, 0.6 and 0.8. We use a soft equation of state with
standard energy-dependent Cugon cross section.
\par

\begin{figure}[!t]
\centering
 \vskip 1cm
\includegraphics[angle=0,width=12cm]{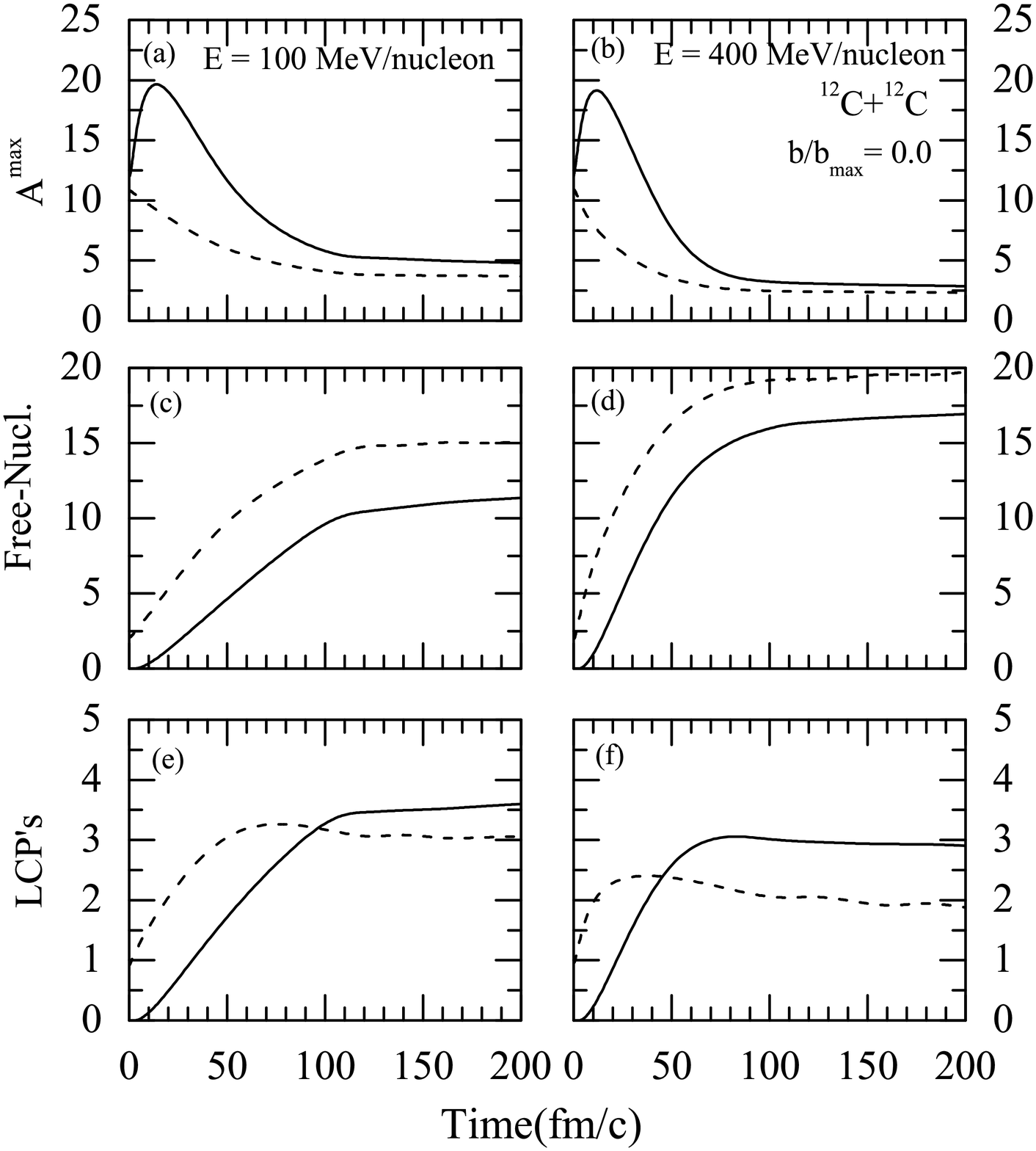}
 \vskip -0cm \caption{ The time evolution of A$^{max}$, free nucleons and LCPs for the reaction of $^{12}$C+$^{12}$C at incident energy of
 100 (left panels) and 400
MeV/nucleon (right) with MST and MSTP methods,
respectively.}\label{fig1}
\end{figure}

\begin{figure}[!t]
\centering \vskip 1cm
\includegraphics[angle=0,width=12cm]{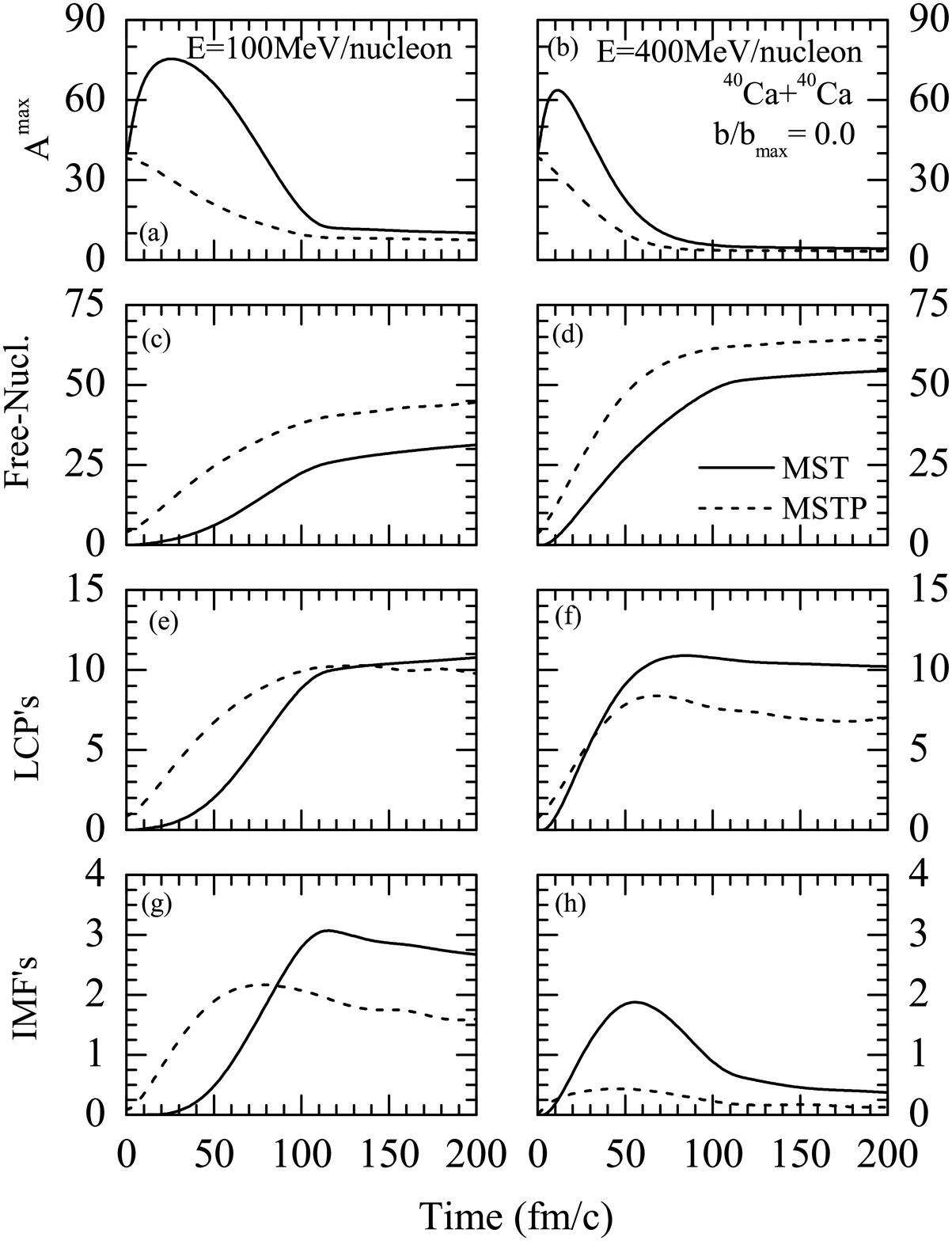}
\vskip -0cm \caption{Same as Fig. 1 but for the reaction of
$^{40}$Ca+$^{40}$Ca.}\label{fig2}
\end{figure}

In Figure 1, we display the time evolution of A$^{max}$[(a),(b)],
free nucleons [(c),(d)] and LCPs(2$\leq$A$\leq$4) [(e),(f)] for
the reactions of $^{12}$C+$^{12}$C at 100 (left panels) and 400
(right) MeV/nucleon. Solid lines indicate the results of MST
method whereas dashed lines represent the results of MSTP method.
The heaviest fragment A$^{max}$ follows different time evolution
in MSTP as compared to MST method. In MST, we have a single big
fragment whereas momentum cut gives two distinct fragments which
shows realistic picture.
\par
In Figure 1(c) and 1(d), we display the time evolution of free
nucleons. We see that for both the energies, MSTP method yields
more free nucleons compared to MST method. There is also a delayed
emission of nucleons in MST because of no restrictions being
imposed. This delayed emission of free nucleons in MST method
taken place because of the fact that till 30 fm/c, we have a
single big fragment in MST method (See Figure 1(a),(b)). The
fragments saturate earlier in MSTP than MST as predicted in Ref.
\cite{kumar1}.
 \par

 \begin{figure}[!t]
\centering \vskip 1cm
\includegraphics[angle=0,width=12cm]{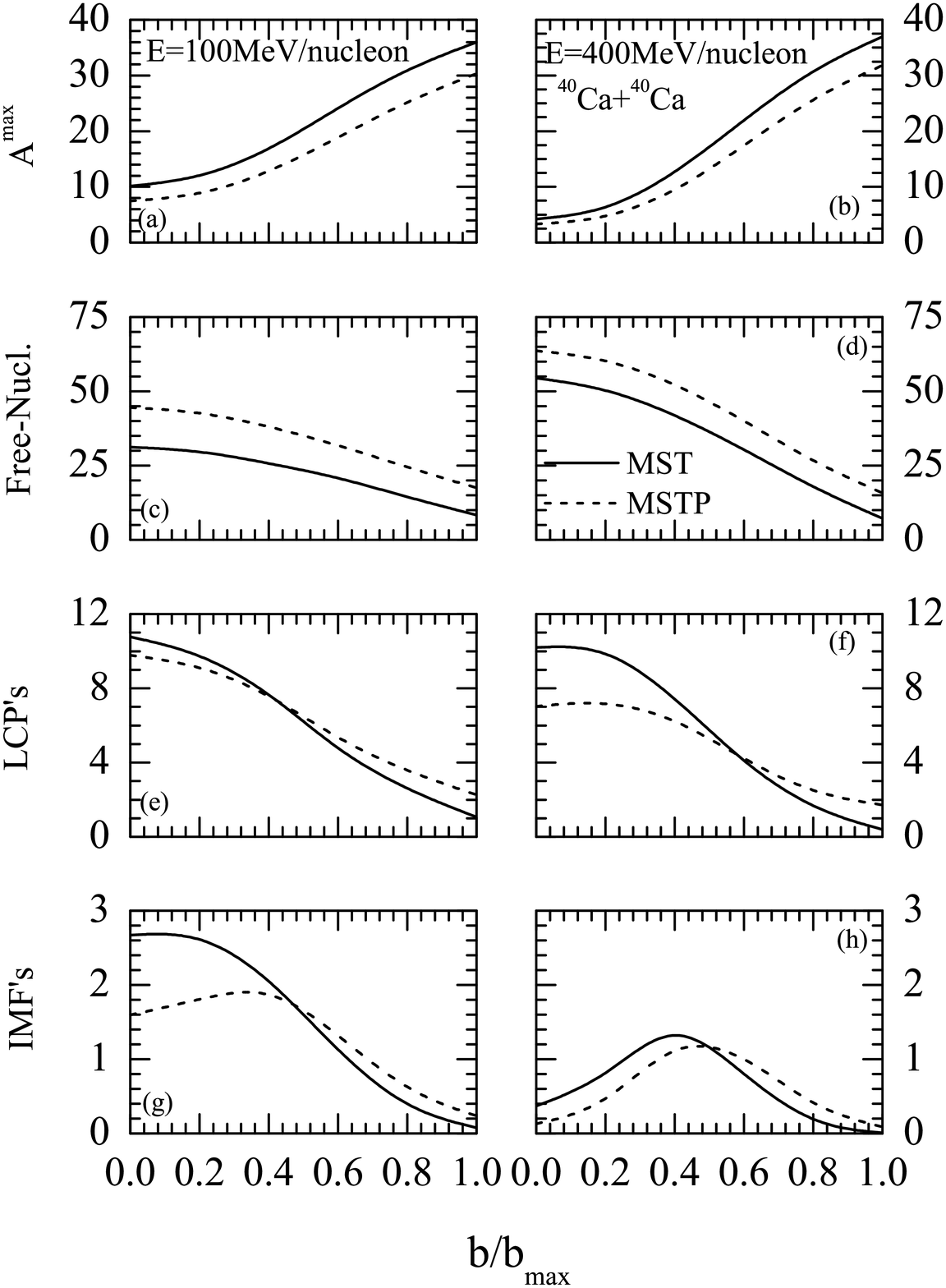}
\vskip -0cm \caption{The impact parameter dependence of A$^{max}$,
free nucleons, LCPs and IMFs for the reaction of
$^{40}$Ca+$^{40}$Ca at 100 (left panels) and 400 (right)
MeV/nucleon with MST and MSTP methods.}\label{fig3}
\end{figure}

In figure 1 [(e),(f)], we display the time evolution of LCPs. We
see that MST yields more LCPs. The difference between MST and MSTP
method increases at 400 Mev/nucleon signifying significant role of
momentum correlations at higher incident energies.
\par

\begin{figure}[!t] \centering
 \vskip 1cm
\includegraphics[angle=0,width=12cm]{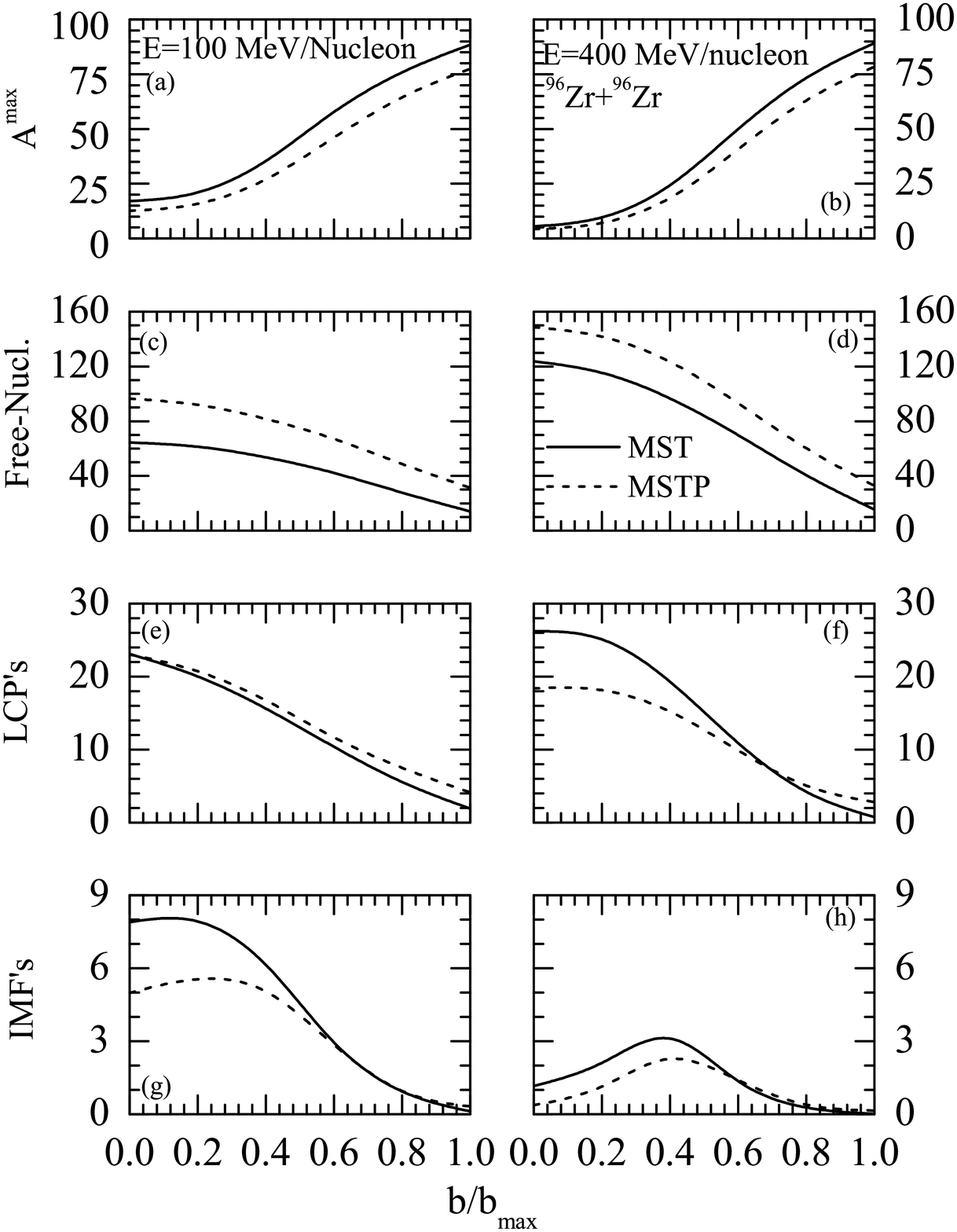}
 \vskip -0cm \caption{ Same as Fig. 3 but for the reaction of
$^{96}$Zr+$^{96}$Zr.}\label{fig5}
\end{figure}

\begin{figure}[!t]
\centering
 \vskip 1cm
\includegraphics[angle=0,width=12cm]{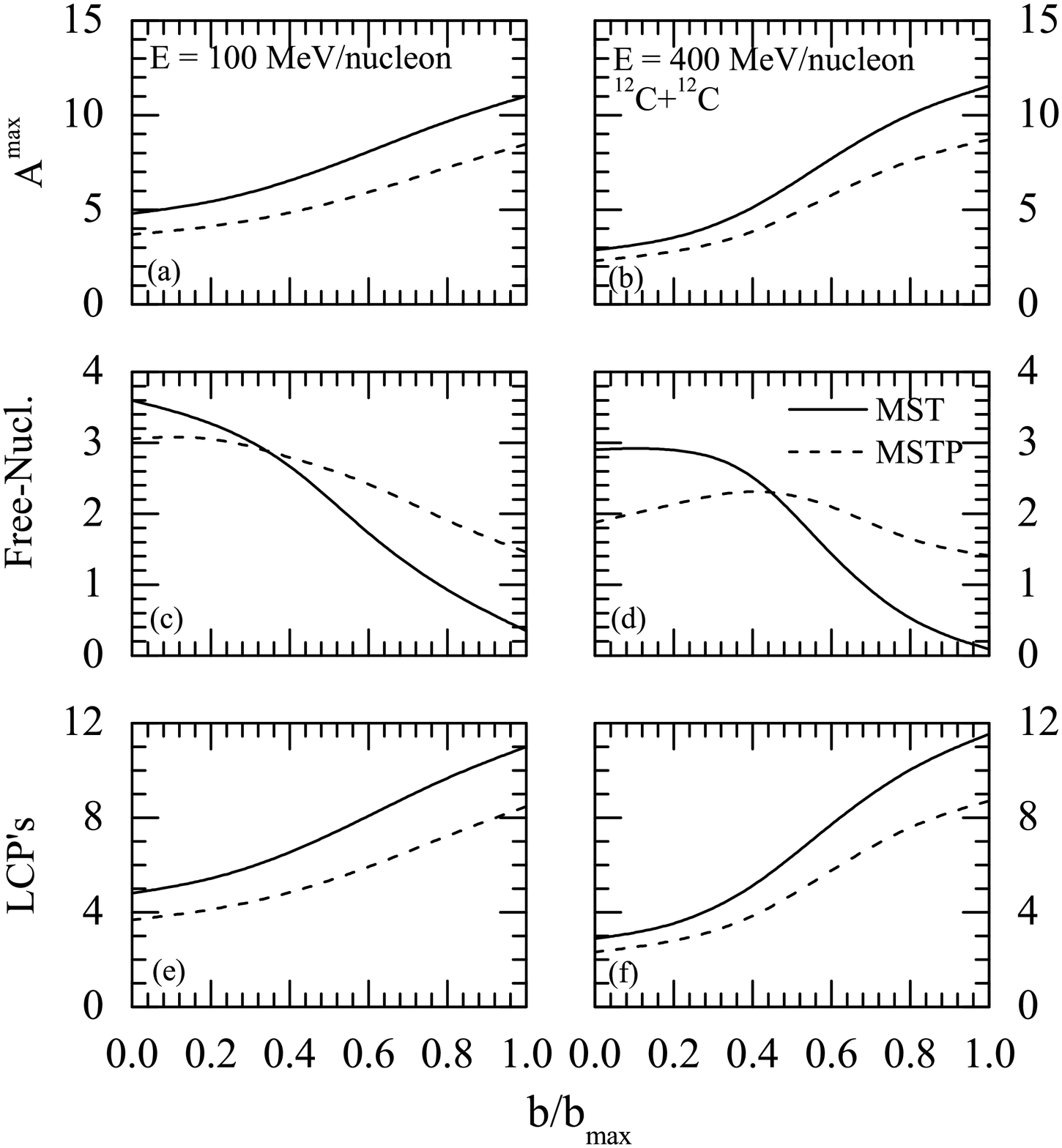}
 \vskip -0cm \caption{ Same as Fig. 3 but for the reaction of
$^{12}$C+$^{12}$C.}\label{fig6}
\end{figure}

In figure 2, we display the time evolution of A$^{max}$, free
nucleons, LCPs and IMFs for the reactions of $^{40}$Ca+$^{40}$Ca
respectively at 100 (left panel) and 400 (right) MeV/nucleon. We
see that A$^{max}$ and free nucleons follow similar behavior as
reported for the reactions of $^{12}$C+$^{12}$C. The emission of
free nucleons enhances with the cut and thus reducing the number
of LCPs and IMFs. Similar effects also seen for the reaction of
$^{96}$Zr+$^{96}$Zr.
\par

In figures 3 and 4, we display the impact parameter dependence of
A$^{max}$, free nucleons, LCPs, and IMFs for the reaction of
$^{40}$Ca+$^{40}$Ca and $^{96}$Zr+$^{96}$Zr, respectively, at
100(left panel) and 400 (right) MeV/nucleon. From both figures we
see that A$^{max}$ rises with impact parameter for both methods
uniformly. The difference increases with impact parameter. This
happens because of the fact that we have a bigger spectator matter
(from where A$^{max}$ generates) at peripheral collisions
geometry. The number of free nucleons decreases with increase in
impact parameter for both methods.
\par

\begin{figure}[!t]
\centering
 \vskip 1cm
\includegraphics[angle=0,width=12cm]{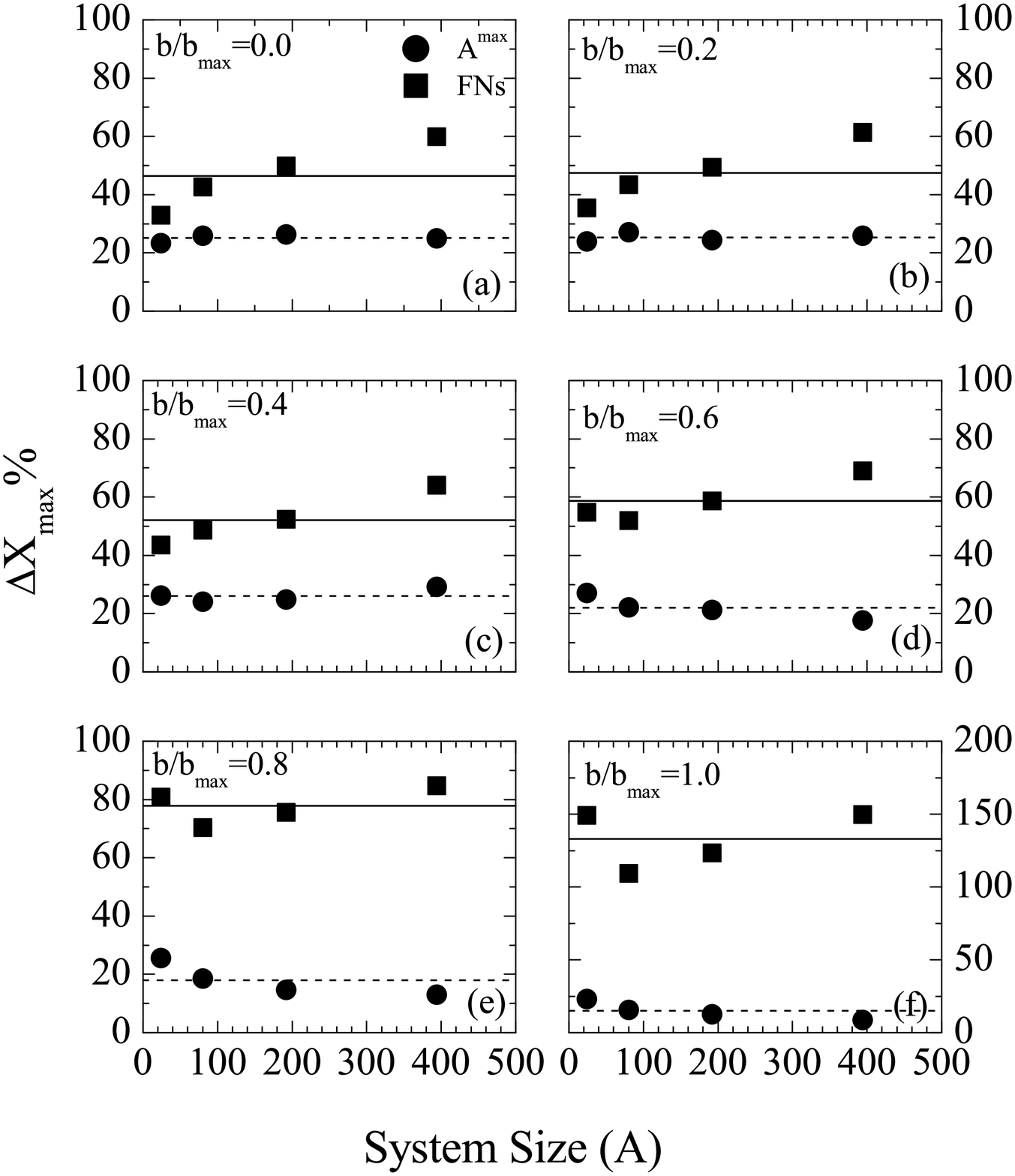}
 \vskip -0cm \caption{ The system size dependence of the percentage difference of A$^{max}$
  and free nucleons between MST and MSTP at various impact parameters. Lines represent the average values.}\label{fig6}
\end{figure}

\begin{figure}[!t]
\centering
 \vskip 1cm
\includegraphics[angle=0,width=12cm]{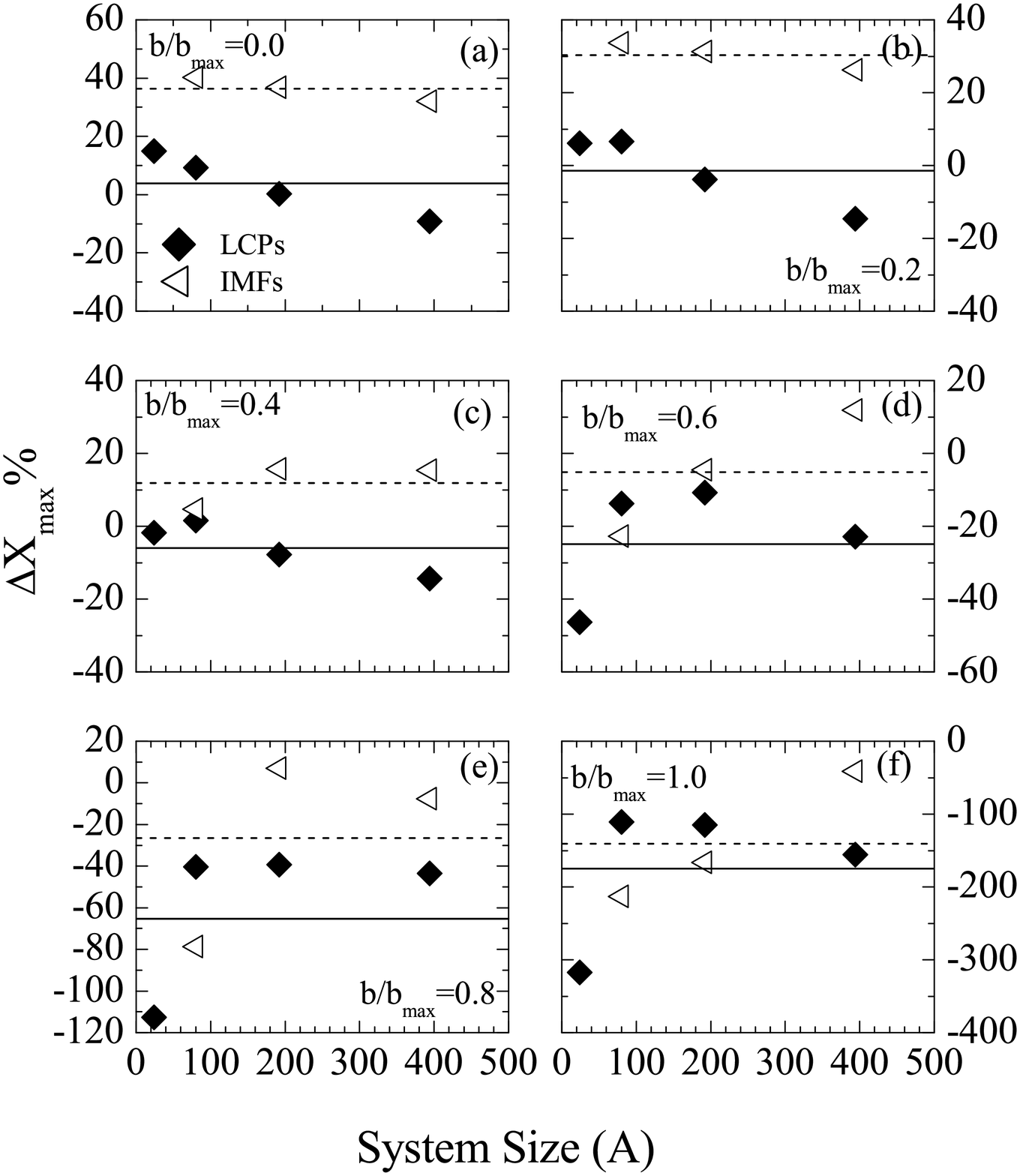}
 \vskip -0cm \caption{ Same as Fig. 6, but for the LCPs and IMFs.}\label{fig6}
\end{figure}

\par
The emission of fragments shows that in contrast to central
collisions, peripheral collisions does not show drastic changes
with method. This happens due to fact that with increase in
colliding geometry, the fragments are the reminants of either
projectile or target, therefore breaking mechanisms are almost
bound and therefore MSTP methods does not give different results.
\par
In figure 5, we display the impact parameter dependence of
A$^{max}$, free nucleons, and LCPs for the reaction of
$^{12}$C+$^{12}$C at 100 (left panel) and 400 (right) MeV/nucleon.
We see that in this particular case, the effect of momentum cut on
the fragment production enhances with impact parameter (see figure
6(e) and (f)) which is quite different compared to earlier
figures. This is because the spectator matter even at peripheral
geometries will be very less in such a lighter system and so the
fragments are emitted mostly from the participant region, where
they are unstable and hence momentum cut plays a role at such
geometries.
\par
\par
In figure 6, we display the percentage difference [$(\Delta
X($\%$) =(X_{(mst)}-X_{(mstp)}/X_{mst}) \times100)$]. We display
the system size dependence of the percentage difference of the
quantities A$^{max}$ and free nucleons at $\hat{b}$= 0.0, 0.2,
0.4, 0.6 and 0.8. We see that percentage difference of A$^{max}$
(circles) is almost constant and remains independent of system
size at central and semicentral colliding geometry where it is
more for lighter systems at peripheral colliding geometries.
Similar behavior is also observed for the free nucleons.
\par
In figure 7, we display the system size dependence of the
percentage difference of LCPs and IMFs at various impact
parameters. From figure we see that in central collision,
$\Delta$IMF\% is almost independent of the system size whereas at
peripheral colliding geometries it increases with system mass. The
average difference is 22, 69, -45, and -16 for A$^{max}$, free
nucleons, LCPs, and IMFs, respectively.

\section{Summary}
 Using quantum molecular dynamic model,we studied the role of
 momentum correlations in fragmentation. This was achieved by
 imposing cut in momentum space during the process of clusterization. We find that this cut yields significant
 difference in the multifragmentations of system at all colliding
 geometries.

\section{Acknowledgement}
This work is done under the supervision of Dr. Rajeev K. Puri,
Department of Physics, Panjab University, Chandigarh, India. This
work has been supported by a grant from Centre of Scientific and
Industrial Research (CSIR), Govt. of India.

\end{document}